\documentclass[12pt]{iopart}

\usepackage{graphicx}
\usepackage{amssymb}

\begin{document}

\title[Algorithm for isometric embedding of
2-surface metrics in 3 dimensional flat space]
{New efficient algorithm for the isometric embedding of
2-surface metrics in 3 dimensional Euclidean space}

\author{Wolfgang Tichy$^{1}$, Jonathan R. McDonald$^{2}$,
        Warner A. Miller$^{1,2}$}

\address{$^{1}$Department of Physics, Florida Atlantic University,
         Boca Raton, FL 33431, USA}
\address{$^{2}$Department of Mathematics, Harvard University,
         Cambridge, MA 02138, USA}


%
\newcommand\be{\begin{equation}}
\newcommand\ba{\begin{eqnarray}}

\newcommand\ee{\end{equation}}
\newcommand\ea{\end{eqnarray}}
\newcommand\p{{\partial}}
\newcommand\remove{{{\bf{THIS FIG. OR EQS. COULD BE REMOVED}}}}
%

\begin{abstract}

We present a new numerical method for the isometric embedding of 2-geometries
specified by their 2-metrics in three dimensional Euclidean space. Our
approach is to directly solve the fundamental embedding equation supplemented
by six conditions that fix translations and rotations of the embedded
surface. This set of equations is discretized by means of a pseudospectral
collocation point method. The resulting nonlinear system of equations are
then solved by a Newton-Raphson scheme. We explain our numerical 
algorithm in detail.
By studying several examples we show that our method converges provided we
start the Newton-Raphson scheme from a suitable initial guess. Our novel
method is very efficient for smooth 2-metrics.

\end{abstract}



\maketitle

\section{Introduction}

The famous Weyl problem~\cite{Weyl1916} is the question whether each
positive definite 2-metric $g_{AB}$ with positive intrinsic Gauss curvature
given on the unit sphere can be realized as the metric on some 2-dimensional
surface embedded in three dimensional Euclidean space
$\mathbb{R}^3$. This realization is referred to as
isometric embedding. Important answers to this question were given first by
Lewy~\cite{Lewy1938} and later by Alexandrov and
Pogorelov~\cite{Aleksandrov1941,Pogorelov1949}
and independently by Nirenberg~\cite{Nirenberg1953} in his groundbreaking
work. The latest answer due to Heinz~\cite{Heinz1962}
is in the affirmative if the metric is three times differentiable. Note,
however, that these answers are complex existence proofs which cannot easily
be used to compute the embedded surface. In order to explicitly
construct this surface we have to find a set of equations that can be solved
at least numerically for any given metric.

The explicit solution of the problem is of interest for example if we want
to visualize black hole horizons. For example current numerical simulations
of black holes routinely compute apparent 
horizons~\cite{Lehner2001,ThornburgLRR}. These
horizons are usually found in particular coordinates that are convenient for
the stability of the simulations, but horizons plotted in these coordinates
can be misleading as deformations can simply be coordinate effects. However,
since the intrinsic horizon metric is known in such simulations its isometric
embedding in $\mathbb{R}^3$ will show the true horizon shape.

Another reason for renewed interest in the isometric embedding problem is
the quest for a quasi-local definition of the energy that is enclosed inside
a specified 2-dimensional surface. A recent definition for this energy by
Wang and Yau~\cite{WangYau2009a,WangYau2009b} 
seems to be particularly promising. It is given in Definition
5.2 of~\cite{WangYau2009b} and involves a minimization. The minimum
is found by solving Eq.~(6.6) of~\cite{WangYau2009b}.
The latter equation is elliptic and contains the extrinsic curvature of a
2-surface that has to be embedded into $\mathbb{R}^3$. Thus in order to find
the energy one has to solve an isometric embedding problem as an
intermediate step in the calculation. Since elliptic problems are well
studied, the hardest part is to find an efficient algorithm that performs
this isometric embedding so that the elliptic equation can be solved and
used to find the minimum needed for the definition of the energy.

The starting point for the embedding problem is simple and given by
\begin{equation}
\label{IE_r_eqn}
  (\partial_A r^x)(\partial_B r^x) + (\partial_A r^y)(\partial_B r^y)
+ (\partial_A r^z)(\partial_B r^z)  = g_{AB} ,
\end{equation}
where $(r^x(u),r^y(u),r^z(u))$ is the vector in $\mathbb{R}^3$ that
describes the embedded surface as a function of the coordinates $u^A$ (with
$A=1,2$) of the given metric on the unit sphere. Since the metric $g_{AB}$
is symmetric Eq.~(\ref{IE_r_eqn}) really consists of three equations for the
unknown functions $r^x,r^y,r^z$. From geometric considerations it is obvious
that the embedded surface is unique only up to arbitrary rigid rotations,
translations and reflections. In addition, Eq.~(\ref{IE_r_eqn}) is not of
a well known type such as e.g. hyperbolic, parabolic or elliptic equations.
Thus standard methods do not necessarily apply. 
According to Khuri, little information has been obtained by studying it     
directly~\cite{Khuri2009}.
For this reason Eq.~(\ref{IE_r_eqn}) is usually reformulated.

One common approach to reformulating Eq.~(\ref{IE_r_eqn}) is to 
multiply it by $du^A du^B$ with the result
\begin{equation} 
(dr^x)^2 + (dr^y)^2 = [g_{AB} - (\partial_A r^z)(\partial_B r^z)]du^A du^B .
\end{equation} 
Since the left hand side is obviously a flat metric, the metric on the right
must also be flat. Setting its intrinsic scalar curvature to zero 
leads to an equation for $r^z$ given by
\begin{equation}
\det(\nabla_A\nabla_B r^z) = K \det(g_{AB}) (1-|\nabla r^z|^2) ,
\end{equation} 
where the Gauss curvature $K=R/2$ is obtained from the Ricci Scalar $R$
of $g_{AB}$.
This equation is called the Darboux equation and contains at most second
derivatives of $r^z$ (because all third derivative terms cancel). The
Darboux equation is of Monge-Ampere type. It is elliptic if $K$ is positive
and hyperbolic if $K$ is negative. Once $r^z$ is known $r^x$ and $r^y$ can
be directly integrated. This approach has been used~\cite{Romano95}
to study Misner initial data for the collision of black holes.

It is also possible to derive a Darboux equation for 
\begin{equation}
\rho := [(r^x)^2 + (r^y)^2 + (r^z)^2]/2 .
\end{equation} 
It reads~\cite{HanHong2006}
\begin{equation}
\label{Darboux_rho}
\det(\nabla_A\nabla_B\rho-g_{AB}) = K \det(g_{AB}) (2\rho-|\nabla\rho|^2) .
\end{equation} 
This formulation has the advantage that we can obtain the extrinsic
curvature (see Eq.~(\ref{extrinsic_curv}))
of the embedded surface directly from $\rho$ without having to
first compute $(r^x, r^y, r^z)$. The disadvantage of the Darboux equation
for $\rho$ is that it is not known what extra conditions we have to impose
on $\rho$ to ensure a unique solution. For example a translation
of $(r^x,r^y,r^z)$ would change the value of $\rho$ as well.
Thus for any solution $\rho$ of Eq.~(\ref{Darboux_rho})) there are
infinitely many nearby solutions that can be generated by infinitesimal
translations. 
Without any extra conditions on $\rho$ this approach is not particularly
suitable for numerical calculations, because most iterative numerical
methods will fail to converge under such conditions.

Another approach~\cite{Nollert98} uses three dimensional wireframes whose
edge lengths are chosen such that they correspond to the lengths between
points computed with the metric $g_{AB}$. In this approach a system of
equations for the points in the wire frame is solved directly.
The approach has been criticized in~\cite{Bondarescu02} for allowing
multiple solutions that are not all smooth. For example if the embedded
surface is supposed to be a sphere one could obtain a wire frame that
corresponds to a sphere whose top third has been inverted, since that would
not change the distances of any neighboring points.
 
Yet another approach~\cite{Bondarescu02} is based on expanding the embedded
surface in spherical harmonics and minimizing the differences between its
metric and $g_{AB}$. As with any minimization scheme there is the danger
that the algorithm gets stuck in a local minimum that does not correspond to
the sought after global minimum where both metrics agree. For this reason
this approach is computationally intensive.

A more recent approach for numerically solving the embedding problem has
been given by Jasiulek and Korzy\'nski~\cite{Jasiulek:2011np}. It uses Ricci
flow to find conformal relations between the original metric, the round
sphere metric and all intermediate metrics. The round sphere metric is
embedded in $\mathbb{R}^3$ and used as the starting point for an embedding
flow back to the original metric. To step from one surface to another,
Eq.~(\ref{IE_r_eqn}) is linearized and solved for the change in the vector
$(r^x,r^y,r^z)$. This method requires an inversion of the extrinsic
curvature tensor at each step therefore limiting its application to
strictly positive scalar curvature surfaces.

Another recent algorithm called AIM~\cite{Ray2014inprogress} starts from a
triangulation of $g_{AB}$. It finds an approximation to the unique Alexandrov
polyhedron for a given triangulated surface with pointwise convex polyhedral
metric by using combinatorial Ricci flow followed by an adiabatic pullback
with annealing. The main advantage of AIM is that it does not require an
initial guess for the embedded surface. The resulting polyhedra 
do not have inverted regions 
unlike the wireframes in~\cite{Nollert98}. While AIM is computationally
demanding, it could provide a good initial guess for other algorithms that
start from an initial guess for the embedded surface.

In this paper we will use a more straight forward approach and simply solve
Eq.~(\ref{IE_r_eqn}) directly. In Sec.~\ref{num_method} we describe our
particular numerical method, followed by some test examples in
Sec.~\ref{results}. We conclude with a discussion of our method in
Sec.~\ref{discussion}.

\section{Numerical method}
\label{num_method}

In this section we present a new numerical method to find the isometric
embedding of a 2-metric $g_{AB}$ on the unit sphere into $\mathbb{R}^3$.
Our approach is to use pseudospectral methods to directly solve
Eq.~(\ref{IE_r_eqn}).
We describe an efficient implementation of this method with the well tested
SGRID code~\cite{Tichy:2006qn,Tichy:2009yr,Tichy:2009zr,Tichy:2012rp}.

\subsection{The pseudo-spectral collocation method}
\label{pseudospec}

In one spatial dimension, spectral methods are based on expansions
\begin{equation}
h(Y) = \sum_{m=0}^{N-1} \tilde{b}_m B_m(Y) 
\end{equation}
of every field $h(Y)$ in terms of suitable basis functions $B_m(Y)$
with coefficients $\tilde{b}_m$. 
Once the coefficients are known it is easy to compute
derivatives of $h(Y)$ from
\begin{equation}
\partial_Y h(Y) = \sum_{m=0}^{N-1} \tilde{b}_m \partial_Y B_m(Y) , 
\end{equation}
since the derivatives of the basis functions are known analytically.

However, instead of directly 
storing and manipulating the coefficients $\tilde{b}_m$
up to some desired order $N-1$ in $m$, we make use of the fact
that (for the basis functions of interest) we can derive $\tilde{b}_m$
from the values of $h(Y)$ at certain collocation points.
If $h(Y)$ is known to have the values $h(Y_j) = h_j$ at the 
collocation points $Y_j$ for $j=0,1,...,N-1$ it is possible
to invert the $N$ equations
\begin{equation}
h_j  = \sum_{m=0}^{N-1}  \tilde{b}_m B_m(Y_j) 
\end{equation}
and to exactly solve for the $N$ coefficients $\tilde{b}_m$ in terms 
of the $h_j$. The location of the different collocation
points depends on the basis functions used. For example, for
Fourier expansions the collocation points have to be equally spaced
in the $Y$-interval considered.
This approach of storing the field's values $h_j$ at the collocation
points is called a pseudo-spectral collocation method.
It has the advantage that non-linear terms such as
$[\partial_Y h(Y_i)]^2$ can be computed from simple multiplications.

The generalization to two dimensions, is straight forward
and can be summarized by
\begin{equation}
h(Y_j,Z_k) 
= \sum_{m,n} \tilde{d}_{mn} B_m(Y_j) C_n(Z_k).
\end{equation}
I.e. we are using basis functions which are products
of functions that depend only on one coordinate. Note that the
coordinates $Y$ and $Z$ need not be Cartesian coordinates.

In this work we have chosen $Y$ and $Z$ to be the standard
spherical coordinates $\theta$ and $\phi$, such that
a point on the unit sphere is given by
$(x,y,z) = (\sin\theta \cos\phi, \sin\theta\sin\phi, \cos\theta)$.
For both angular directions we use a Fourier basis so that
\begin{eqnarray}
\label{thetabasisfunc}
B_m(\theta_j) &=& F_m^{N_{\theta}}(\theta_j) \\
\label{phibasisfunc}
C_n(\phi_k) &=&  F_n^{N_{\phi}}(\phi_k) ,
\end{eqnarray}
where
\begin{equation}
\label{realFourier}
F_l^N(\varphi) 
=  \left\{
\begin{array}{ll}
1/N				& \mbox{if $l=0$}     \\
2\cos(\frac{l+1}{2} \varphi)/N	& \mbox{if $l$ is odd}     \\
2\sin(\frac{l}{2} \varphi)/N	& \mbox{if $l$ is even}     \\
\cos(\frac{N}{2} \varphi)/N	& \mbox{if $l=N-1$ and $N$ is even.}     \\
\end{array}
\right.
\end{equation}
The collocation points are chosen to be
\begin{eqnarray}
\label{theta_j}
\theta_j &=& 
\frac{2\pi j}{N_{\theta}} + \frac{2\pi}{[3-(-1)^{N_{\theta}}] N_{\theta}}, \\
\label{phi_k}
\phi_k &=&
\frac{2\pi k}{N_{\phi}} ,
\end{eqnarray}
where
\begin{eqnarray}
j &=& 0,1,...,N_{\theta} -1 \\
k &=& 0,1,...,N_{\phi} -1 .
\end{eqnarray}
This choice ensures that there are no collocation points at the 
coordinate singularities located at $\theta=0$ and $\theta=\pi$.
Notice that, since both $\theta_j$ and $\phi_k$ are between 0 and $2\pi$,
we have a double covering of the entire domain. 
This double covering is necessary to ensure the periodicity required
for Fourier expansions in both angles.

Any function $h(\theta,\phi)$ can then be expressed in this basis as 
\begin{equation}
\label{SphericalDF0}
h_{j,k} = h(\theta_j,\phi_k)
= \sum_{m=0}^{N_{\theta}-1} \sum_{n=0}^{N_{\phi}-1}
\tilde{d}_{mn} B_m(\theta_j) C_n(\phi_k) . 
\end{equation}
However, in our code we never really uses this expansion to
compute all the coefficients $\tilde{d}_{mn}$. Rather, we only
ever expand in one direction and instead use
\begin{eqnarray}
\label{SphericalDF_theta}
h_{j,k}
&=& \sum_{m=0}^{N_{\theta}-1} \tilde{b}_{l}(\phi_k) B_m(\theta_j) , \\
\label{SphericalDF_phi}
h_{j,k}
&=& \sum_{n=0}^{N_{\phi}-1} \tilde{c}_{l}(\theta_j) C_n(\phi_k) ,
\end{eqnarray}
to compute the coefficients
$\tilde{b}_{l}(\phi_k)$ or $\tilde{c}_{l}(\theta_j)$
along a line in the $\theta$- or $\phi$-direction. 
This suffices to compute partial derivatives with respect to our coordinates
$\theta$ and $\phi$.

\subsection{Solving the isometric embedding equation}

In order to solve the isometric embedding Eq.~(\ref{IE_r_eqn}) we represent
each component of the vector $(r^x,r^y,r^z)$ by its 
values $(r^x_{j,k},r^y_{j,k},r^z_{j,k})$ at the
collocation points of Eqs.~(\ref{theta_j}) and (\ref{phi_k}). This results in
$3 N_{\theta} N_{\phi}$ unknowns that have to be determined. Since
Eq.~(\ref{IE_r_eqn}) is symmetric in $A$ and $B$ we have 3 equations per
collocation point which results in $3 N_{\theta} N_{\phi}$ equations in
total. However, as already mentioned in the introduction,
Eq.~(\ref{IE_r_eqn}) has a unique solution only up to arbitrary rigid
rotations, translations and reflections. This means that not all these 
equations are independent. In order to fix the position and orientation
of the embedded surface we replace six of the equations by the following six
conditions. 
The first three conditions replace the equation for $g_{\theta\phi}$
in Eq.~(\ref{IE_r_eqn})
at the coordinates with indices $(j,k)=(0,0)$, $(j,k)=([N_{\theta}/4],0)$ and
$(j,k)=([(N_{\theta}-1)/2],0)$ by
\begin{equation}
\label{ry_cond}
r^y_{0,0} = 
r^y_{[N_{\theta}/4],0} = 
r^y_{[(N_{\theta}-1)/2],0} = 0 .
\end{equation}
Note that $[x]$ here means: round $x$ to the closest integer below or at
$x$. The conditions (\ref{ry_cond}) ensure that three points with $\phi=0$
will have $r^y=0$. Once $r^y$ is fixed in this way, the embedded surface can
still be translated in the $z$- and $x$-directions and also rotated about
the $y$-axis. The two translations are fixed by demanding
\begin{eqnarray}
\label{transl_fix}
r^z_{0,0} 		&=& -r^z_{[(N_{\theta}-1)/2],0} \\
r^x_{[N_{\theta}/4],0}	&=& -r^x_{[N_{\theta}/4],[N_{\phi}/2]} .
\end{eqnarray}
These two conditions replace the equation for $g_{\phi\phi}$ at the point
$(j,k)=(0,0)$ and the equation for $g_{\theta\theta}$ at the point   
$(j,k)=([N_{\theta}/4],0)$. They ensure that the embedded surface
is centered near $x=y=z=0$.
The remaining rotation freedom about the $y$-axis is fixed by replacing the
equation for $g_{\theta\theta}$ at the point with index $(j,k)=(0,0)$ by
\begin{equation}
\label{roty_fix}
r^x(\theta=0,\phi=0) = 0 .
\end{equation}
Note that $r^x(\theta=0,\phi=0)$ has to be computed by spectral
interpolation because there is no collocation point at $\theta=0$.
This last condition fixes the orientation of the embedded surface such that
$r^x=0$ at the point $\theta=\phi=0$.

The particular components and collocation points at which we replace
Eq.~(\ref{IE_r_eqn}) by any six conditions that fix translations and
rotations do not matter in principle. Our particular choices above are
mainly motivated by trying to obtain simple expressions for the conditions.
Choosing points near the poles and the equator of the $\theta,\phi$
coordinate system is one way of obtaining such simple equations, but other
choices are possible.

If we take all these conditions into account we obtain 
$N = 3 N_{\theta} N_{\phi}$ non-linear equations of the form
\begin{equation}
f_m (w) = 0, \ \  m=1,2,...,N
\end{equation}
for the $N$ unknown $(r^x,r^y,r^z)$ at all collocation points which make up
the solution vector $w$.
We solve this system of equations by a Newton-Raphson scheme.
In order to solve the linearized equations
\begin{equation}
\label{linearEqs}
\frac{\partial f_m(w)}{\partial w^n } x^n = - f_m (w)
\end{equation}
in each Newton-Raphson step, we note that $f_m (w)$ contains
spectral derivatives of $w$ in different directions, so that
the $N\times N$ matrix $\frac{\partial f_m(w)}{\partial w^n }$ is sparse
in the sense that it contains about 95\% zeros.
We use the sparse matrix solver 
UMFPACK~\cite{Davis-Duff-1997-UMFPACK,Davis-Duff_UMFPACK_1999,
Davis_UMFPACK_V4.3_2004,Davis_UMFPACK_2004,umfpack_web}
to numerically solve the linearized Eq.~(\ref{linearEqs}).
This scheme requires an initial guess, for which we simply
use a spherical surface. 

We have observed that the Newton-Raphson scheme only reliably converges if
both $N_{\theta}$ and $N_{\phi}$ are chosen to be odd. The reason for this
is likely related to the fact that only for an odd number of collocation
points we have the same number of sine and cosine functions in the
expansions given by Eqs.~(\ref{thetabasisfunc}), (\ref{phibasisfunc}), and
(\ref{realFourier}). For an even number of collocation points, there is only
a cosine function at the highest wavenumber. So the derivative of this term
(which would be a sine function) cannot be represented with our expansion,
which in turn means that the highest coefficient never enters our system of
equations and thus cannot be determined.

\section{Results}
\label{results}

In this section we will use our method to numerically find the embedding
of several 2-metrics. In order to compare with analytically known results it
is useful to compute the extrinsic curvature~\cite{HanHong2006}
\begin{equation}
\label{extrinsic_curv}
h_{AB} = 
\frac{\nabla_A \nabla_B \rho - g_{AB}}
     {\sqrt{2\rho - g^{AB} \nabla_A\rho \nabla_B\rho}}
\end{equation}
of the surface embedded in $\mathbb{R}^3$.
We now demonstrate our method for a few simple examples.

\subsection{Ellipsoid}

We first consider the following 2-metric:
\begin{eqnarray}
\label{ellipsoid}
g_{\theta\theta} &=& 
( a^2  \cos^2\phi + b^2 \sin^2\phi ) \cos^2\theta  + c^2 \sin^2\theta
\nonumber \\
g_{\theta\phi} &=& 
( b^2 - a^2 ) \sin\phi \cos\phi \sin\theta \cos\theta
\nonumber \\
g_{\phi\phi} &=& ( b^2 \cos^2\phi + a^2 \sin^2\phi ) \sin^2\theta  
\end{eqnarray}
This is the metric of an ellipsoid in $\mathbb{R}^3$ with an aspect
ratio of $a:b:c$. Thus the embedding is known analytically and given
by
\begin{equation}
(r^x,r^y,r^z) =
( a \cos\phi \sin\theta ,
  b \sin\phi \sin\theta ,
  c\cos\theta )
\end{equation}
and 
\begin{eqnarray}
\label{ana_ellpsoid_curv}
h_{\theta\theta} &=&
-a b c/
           \sqrt{a^2 b^2 \cos^2\theta + c^2 (b^2 \cos^2\phi +
                a^2 \sin^2\phi) \sin^2\theta}
\nonumber \\
h_{\theta\phi} &=& 0
\nonumber \\  
h_{\phi\phi} &=& -a b c \sin^2\theta/
           \sqrt{a^2 b^2 \cos^2\theta + c^2 (b^2 \cos^2\phi +
                a^2 \sin^2\phi) \sin^2\theta}
\end{eqnarray}

We have used our numerical method with the metric given in
Eq.~(\ref{ellipsoid}) for $a:b:c=3:2:1$. We start our algorithm from an
initial spherical surface with a radius of $r=\sqrt{2}$. For a resolution of
$N_{\theta} = N_{\phi} = 15$ a solution is found after 5 Newton-Raphson
steps. The radius of the initial surface is not important. Our method
also works with other initial guesses, such as larger radii or 
off-centered spheres.

\begin{figure}
\includegraphics[scale=0.5,clip=true]{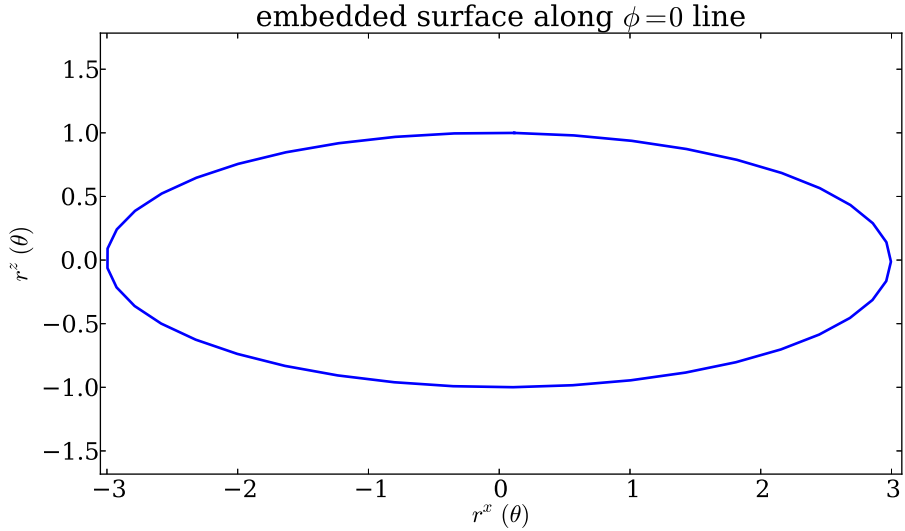}
\caption{\label{rz-rx_ellipsoid}
The embedding of the ellipsoid metric leads again to an ellipsoid. The plot
shows $r^z$ vs. $r^x$ for $\phi=0$.
}
\end{figure}

\begin{figure}
\includegraphics[scale=0.5,clip=true]{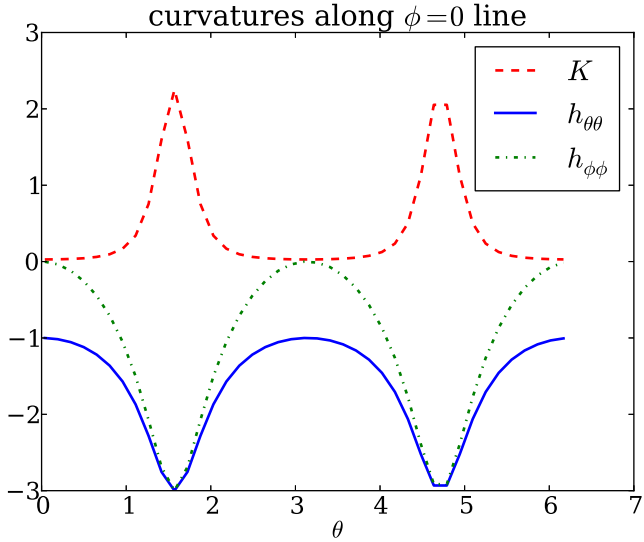}
\caption{\label{K_h_ellipsoid}
The Gauss curvature for the ellipsoid metric with $a:b:c=3:2:1$
drops to 1/36 at $\theta=0$, but is nowhere zero.
Two components of the extrinsic curvature $h_{AB}$ are shown as well.
}
\end{figure}

The embedded surface for $\phi=0$ is shown in
Fig.~\ref{rz-rx_ellipsoid}. We see that our method works as expected
and recovers the ellipsoid from the 2-metric.
The resulting extrinsic curvature is shown in Fig.~\ref{K_h_ellipsoid}
together with the intrinsic Gauss curvature.
The $L^2$-norm of its error
\footnote
{The $L^2$-norm of the error is not normalized. 
For a component like $h_{\theta\theta}$ we compute it from
$\sqrt{\sum_{j,k}
 (h_{\theta\theta}-h_{\theta\theta}^{\tiny\mbox{analytic}})^2
/(N_{\theta}N_{\phi})}$
where we sum over all grid points.
}
is about $10^{-12}$ when we compare with the analytic result
in Eq.~(\ref{ana_ellpsoid_curv}).

\subsection{Deformed ellipsoid}

The next test is to find an embedding for the metric
\begin{eqnarray}
\label{def_ellipsoid}
g_{\theta\theta} &=&
  (b \cos\theta \sin\phi - \frac{1}{2} \cos\theta \sin\theta)^2 +
  (a \cos\phi \cos\theta + \frac{1}{2} \cos\theta \sin\theta)^2 \nonumber \\
& & + (c - \frac{1}{8}\cos\theta)^2 \sin^2\theta
\nonumber \\
g_{\theta\phi} &=& 
\cos\theta (b \cos\phi (b \sin\phi - \frac{1}{2} \sin\theta) -
   a \sin\phi (a \cos\phi + \frac{1}{2} \sin\theta)) \sin\theta
\nonumber \\
g_{\phi\phi} &=&
(b^2 \cos^2\phi + a^2 \sin^2\phi) \sin^2\theta .
\end{eqnarray}
This metric belongs to a deformed ellipsoid. Thus the analytic solution
of the embedding problem is known to be 
\begin{eqnarray}
r^x &=& a \cos\phi \sin\theta + \frac{1}{4}\sin^2\theta  \nonumber \\
r^y &=& b \sin\phi \sin\theta + \frac{1}{4}\cos^2\theta  \nonumber \\
r^z &=& c\cos\theta + \frac{1}{16}\sin^2\theta .
\end{eqnarray}
    
\begin{figure}
\includegraphics[scale=0.5,clip=true]{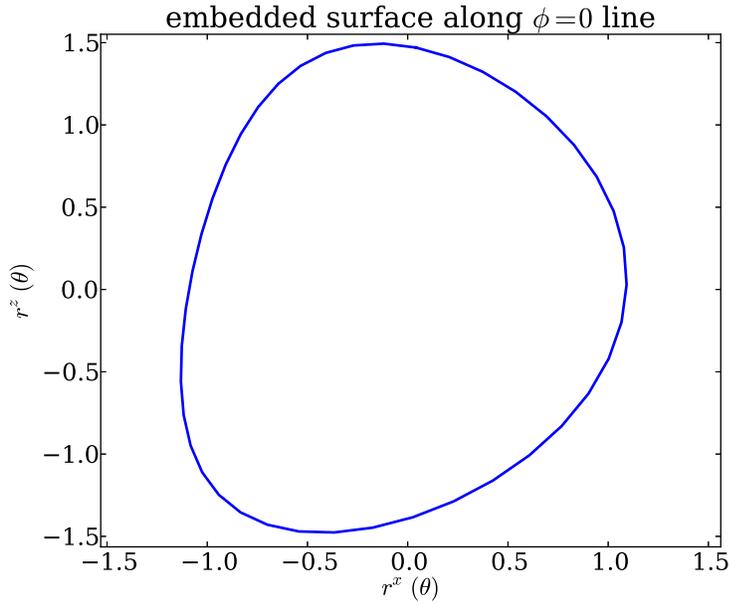}
\caption{\label{rz-rx_deformed}
The shape of the embedded surface for $\phi=0$ when we embedded the 2-metric
of the deformed ellipsoid with $a:b:c=1.1:1.2:1.5$.
}
\end{figure}

\begin{figure}
\includegraphics[scale=0.5,clip=true]{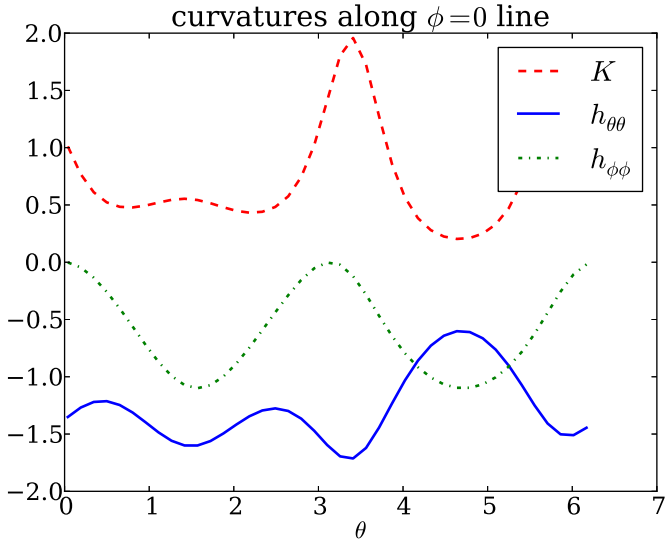}
\caption{\label{K_h_deformed}
The Gauss curvature of the deformed ellipsoid is everywhere positive.
The non-zero components of the extrinsic curvature are shown as well.
}
\end{figure}

Using the metric given in Eq.~(\ref{def_ellipsoid}) for $a:b:c=1.1:1.2:1.5$,
we have started our algorithm from an initial spherical surface with a radius
of $r=1.12546$. For a resolution of $N_{\theta} = N_{\phi} =15 $ a solution
is again found after 5 Newton-Raphson steps and the error is again only of
order $10^{-12}$. Again the initial radius is not important and
$r=1.12546$ is just the exact value we have used to produce the data for
the figures in this subsection.
Other initial radii also work, but may need more Newton-Raphson steps.
For example for $r=100$ we need 11 Newton-Raphson steps.
The embedded surface is shown in Fig.~\ref{rz-rx_deformed}.
The resulting curvatures are shown in Fig.~\ref{K_h_deformed}.
Despite the more irregular shape of the Gauss curvature the algorithm has no
problem finding the correct isometric embedding.

\subsection{Horizon shape of a Kerr black hole}

As already mentioned in the introduction, the isometric embedding of the
metric on a black hole horizon can be useful for visualizing the intrinsic
horizon geometry. Here we consider the horizon of an axisymmetric 
Kerr black hole.
The metric on the horizon in this case is well known. For a black hole of
mass $m$ and spin parameter $a$ it is given by
\begin{eqnarray}
\label{KerrHorizon}
g_{\theta\theta} &=& r_H^2 + a^2 \cos^2\theta  \nonumber \\
g_{\theta\phi}   &=& 0    \nonumber \\
g_{\phi\phi}     &=& 4m^2(2m r_H - a^2)\sin^2\theta/g_{\theta\theta}
\end{eqnarray}
where 
\begin{equation}
r_H = m + \sqrt{m^2 - a^2}
\end{equation}
is the horizon radius and we are using units of $G=c=1$.
Note that a horizon exists only for $a<m$.
\begin{figure}
\includegraphics[scale=0.5,clip=true]{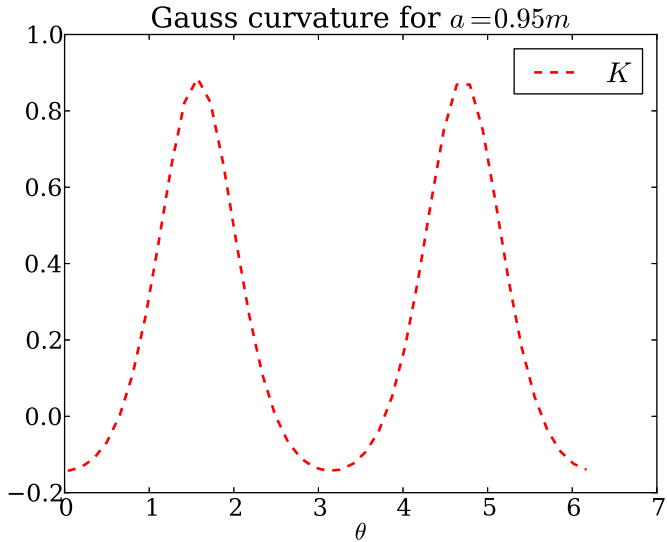}
\caption{\label{K_KerrHorizon_a095}
Gauss curvature of a highly spinning black hole horizon
(in units of $G=c=m=1$) that cannot be embedded in $\mathbb{R}^3$.
}
\end{figure}
Figure~\ref{K_KerrHorizon_a095} shows the Gauss curvature of the horizon
for the case of $a=0.95m$.
As we can see it becomes negative near the poles of the black hole.
This horizon cannot be embedded in
$\mathbb{R}^3$ at all. In fact it is well known that only horizons with 
$a<\sqrt{3}m/2$ can be embedded in 
$\mathbb{R}^3$~\cite{Bondarescu02,Smarr73b}.
Such horizons have $K>0$ everywhere. Thus our algorithm of course can
only work if $a<\sqrt{3}m/2$.
\begin{figure}
\includegraphics[scale=0.5,clip=true]{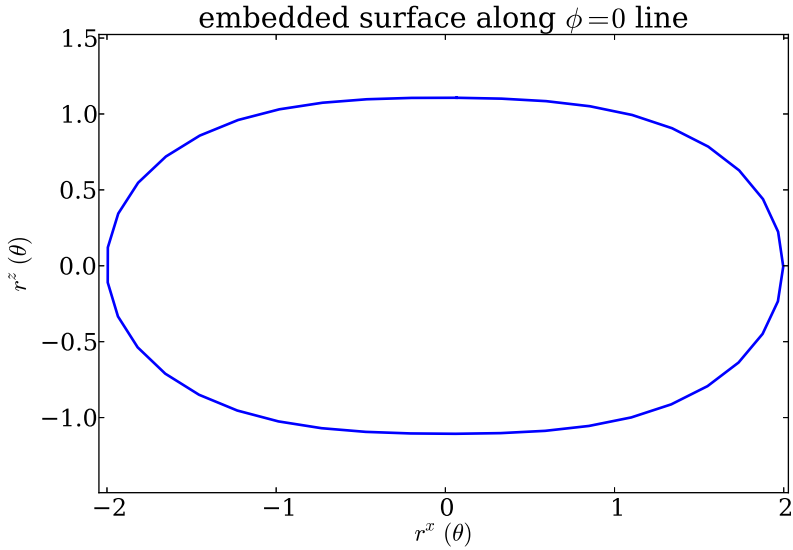}
\caption{\label{rz-rx_KerrHorizon}
Horizon shape (in units of $G=c=m=1$)
of a Kerr black hole with $a=0.86m$. The horizon in this case
is axisymmetric and looks the same for all $\phi$.
}
\end{figure}
For example, for $a=0.86m$ which is less than 1\% from 
the limit our algorithm still works. In this case the embedded surface
(shown in Fig.~\ref{rz-rx_KerrHorizon}) becomes flattened at the poles
with $K$ nearly zero.
\begin{figure}
\includegraphics[scale=0.5,clip=true]{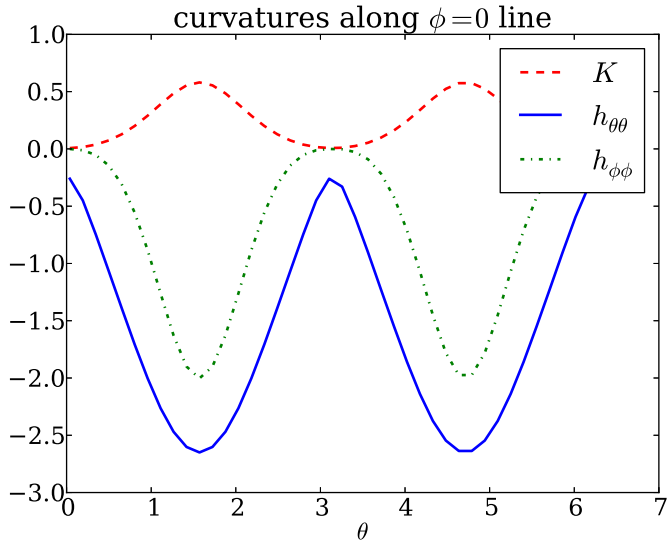}
\caption{\label{K_h_KerrHorizon}
Gauss curvature and extrinsic curvatures for a horizon 
with $a=0.86m$ that is close to the embedding limit of $a<\sqrt{3}m/2$.
$K$ is barely positive at the poles.
}
\end{figure}
The corresponding curvatures are plotted in Fig.~\ref{K_h_KerrHorizon}.

\subsection{Dented sphere}

As we have seen from the example of the Kerr black hole horizon,
an embedding for $K<0$ into $\mathbb{R}^3$ does not always exist.
On the other hand there are plenty of closed surfaces in $\mathbb{R}^3$
that have negative curvature in some places. Imagine for example a sphere
with a dent (with continuous extrinsic curvature). Even though 
the Gauss curvature is expected to be negative around the dent,
the metric on a dented sphere is certainly embeddable in $\mathbb{R}^3$.

In this subsection we consider a particular example of a deformed sphere
that while still axisymmetric has a dent at each pole.
Our particular metric has the form
\begin{eqnarray}
\label{dentedSphere}
g_{\theta\theta} &=& \cos^2\theta + \sin^2\theta [3\cos(2\theta)-1]^2 /4
 \nonumber \\
g_{\theta\phi}   &=& 0
 \nonumber \\
g_{\phi\phi}     &=& \sin^2\theta .
\end{eqnarray}
Its embedding is given by
\begin{eqnarray}
r^x &=& \cos\phi \sin\theta  \nonumber \\
r^y &=& \sin\phi \sin\theta  \nonumber \\
r^z &=& (1+\sin^2\theta) \cos\theta .
\end{eqnarray}

When we use our method with the metric given in Eq.~(\ref{dentedSphere}) and
start our algorithm from an initial spherical surface with radius $r=1$ the
Newton-Raphson scheme does not converge. This problem occurs for all
spherical initial guesses for all radii.
We have not fully investigated if it can be circumvented by a more
sophisticated algorithm. However, we have tried a modified Newton-Raphson
scheme that uses backtracking, i.e. it searches for the optimum step along
the direction found by the linear solver but may take a smaller step if that
reduces the residual error. This algorithm also fails. It eventually gets
stuck because the optimum step becomes a step of length zero.

\begin{figure}
\includegraphics[scale=0.5,clip=true]{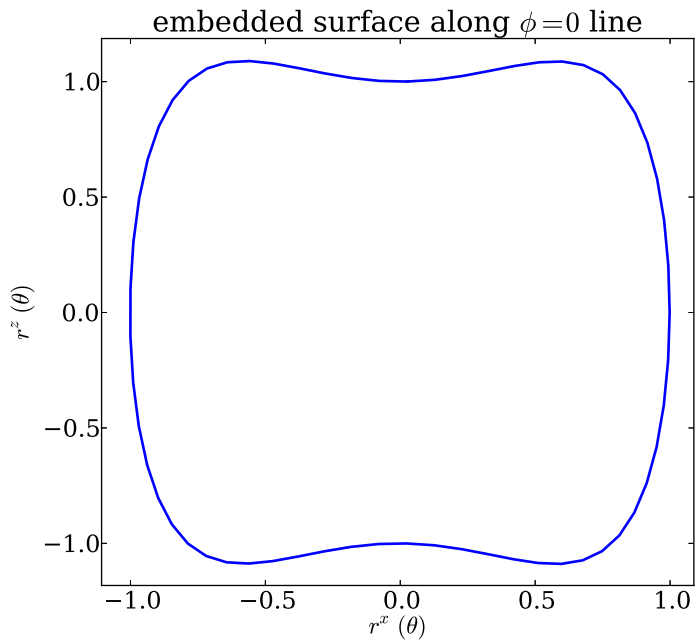}
\caption{\label{rz-rx_dented}
The embedded surface for the case of a metric that comes from a sphere with
dents at the poles. The surface is axisymmetric.
}
\end{figure}

\begin{figure}
\includegraphics[scale=0.5,clip=true]{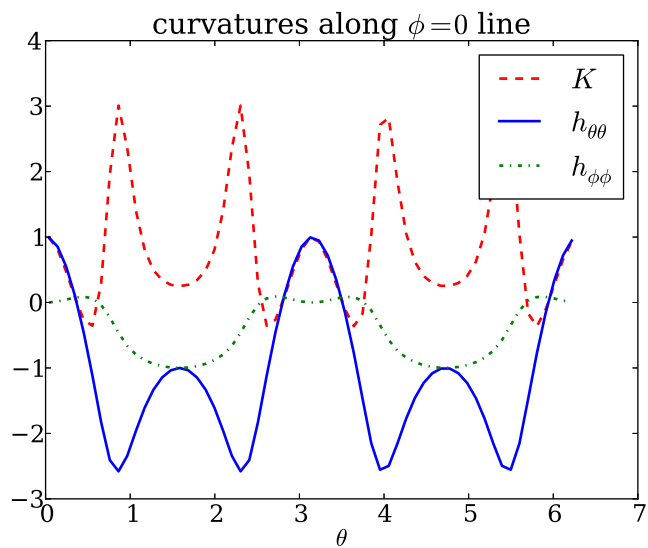}
\caption{\label{K_h_dented}
The Gauss curvature of the dented sphere is not positive definite.
The non-zero components of the extrinsic curvature are shown as well.
}
\end{figure}

On the other hand if we start with 
\begin{eqnarray}
\label{dent_guess}
r^x_{guess} &=& 0.9 \cos\phi \sin\theta  \nonumber \\
r^y_{guess} &=& 0.7 \sin\phi \sin\theta  \nonumber \\
r^z_{guess} &=& 2 (1+\sin^2\theta) \cos\theta
\end{eqnarray}
as initial guess, or something similar, the scheme converges and finds the
correct solution.
The embedded surface found by the scheme in this case is shown in
Fig.~\ref{rz-rx_dented}. We see that this surface has large dents in the
polar regions.
In Fig.~\ref{K_h_dented} we show the resulting curvatures. It is obvious
that there are large regions where $K<0$ around both poles.

The reason why an initial guess similar to Eq.~(\ref{dent_guess}) works,
while a spherical initial guess fails, may be related to the fact that the
guess in Eq.~(\ref{dent_guess}) already has two dents just like the true
embedded surface.

In order to verify our solution we have also conducted a convergence test
for the dented sphere case. Notice, however, that the particular 2-metric
in Eq.~(\ref{dentedSphere}) can be expanded in terms of a finite 
number of sine and
cosine functions. Since our spectral method basically consists of expanding
everything in sine and cosine functions, our truncated expansion
becomes exact as soon as 
we expand to high enough order, i.e. if we use enough collocation points. 
As is evident from Eq.~(\ref{dentedSphere}) the sine and cosine functions
with the highest frequency will have $6\theta$ as an argument. From
Eq.~(\ref{realFourier}) we can thus see that we need at least
$N_{\theta}=13$ to include all these functions in our spectral expansions.
So a convergence plot is expected to show errors that fall with increasing
number of collocation points only for $N_{\theta}<13$.
For $N_{\theta} \ge 13$ the truncation error will be zero and
the residual errors will be dominated by
numerical round off errors due to the use of floating point arithmetic, i.e.
the error should level off at ca. $10^{-12}$. 
\begin{figure}
\includegraphics[scale=0.5,clip=true]{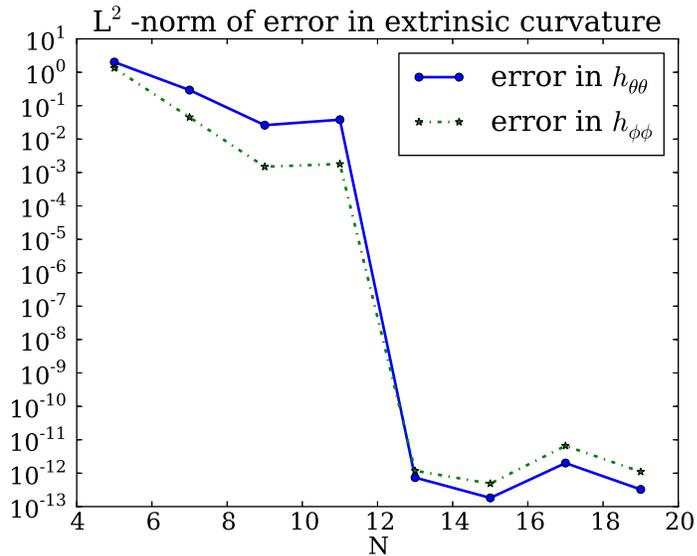}
\caption{\label{h_Err_dented}
The errors in the extrinsic curvature
for different numbers of collocation points $N = N_{\theta} = N_{\phi}$
converge until the spectral method becomes exact for $N \ge 13$. 
}
\end{figure}
In Fig.~\ref{h_Err_dented} we show the errors in the extrinsic curvature
for different numbers of collocation points $N = N_{\theta} = N_{\phi}$.
We see the expected approximately exponential convergence for $N < 13$,
while as anticipated we reach numerical round off for $N \ge 13$.

Similar arguments also apply to all the other examples discussed in the
previous sections. As soon as $N$ is large enough the spectral expansions
become exact for all our examples. Notice, however, that this only happens
due to the simplicity of our examples. For a generic smooth 2-metric the
spectral expansions would not be exact for any finite $N$.

\section{Discussion}
\label{discussion}

The purpose of this paper is to introduce a new numerical method
for the computation of the isometric embedding of 2-metrics in
$\mathbb{R}^3$.
Our method directly solves the embedding Eq.~(\ref{IE_r_eqn}) 
together with 6 conditions 
(given in Eqs.~(\ref{ry_cond}), (\ref{transl_fix}) and (\ref{roty_fix}))
that fix the center and orientation of the embedded surface.
We use a pseudospectral collocation point method to represent the 
derivatives in Eq.~(\ref{IE_r_eqn}). This results in three nonlinear equations per
collocation point for the three components $(r^x,r^y,r^z)$
of the embedding vector. This set of equations is then solved with a
Newton-Raphson scheme.

As we have seen it is important to work with an odd number of collocation
points in both coordinate directions if one uses Fourier expansions. As long
as the given 2-metric has positive Gauss curvature everywhere our method
seems to always converge if we initialize the Newton-Raphson scheme with a
spherical surface as initial guess. For cases where the Gauss curvature is
negative in some regions, but an isometric embedding is known to exist, our
method can fail if we use a spherical surface as initial guess for the
Newton-Raphson scheme. Nevertheless the scheme can be made to converge by
choosing a better initial guess.

Our algorithm is very efficient. On a laptop computer for $N_{\theta} =
N_{\phi} = 15$ it only takes about half a second to find the isometric
embedding of the deformed metric of Eq.~(\ref{def_ellipsoid}) shown in
Fig.~\ref{rz-rx_deformed}. For comparable accuracies this is orders of
magnitude faster than the AIM algorithm~\cite{Ray2014inprogress}. However,
since most other previous methods have used different examples and since the
papers describing them do not give exact timing information we cannot
directly compare. Nevertheless, our method has the advantage of not allowing
multiple solutions, so that it does not suffer from the drawbacks that can
cause problems for wireframe approach~\cite{Nollert98}. We also expect it to
be faster than the minimization scheme discussed in~\cite{Bondarescu02}.

\ack
It is a pleasure to thank Shannon Ray
for useful discussions about isometric embeddings,
and Shing-Tung Yau for motivating this research.
This work was supported by NSF grant PHY-1305387
and by a grant from the Air Force Research Laboratory (AFRL/RITA),
Grant \#FA8750-11-2-0089. 
WAM would like to acknowledge support from the VFRP program through Griffiss
Institute.






\vskip 1cm


\bibliographystyle{unsrt}
\bibliography{references}

\end{document}